\documentclass[12pt]{article}
\usepackage{epsf}
\usepackage{graphicx}
\usepackage{lscape}
\usepackage{amssymb}
\usepackage{amsmath,amsfonts}
\usepackage{pazh}
\tightenlines

\def\ΧΆ{$\pm$}
\def\*{$^{*}$}

\def\Œ³ŒÀ{\mbox{\<<Œ³ŒàŒÐŒÝŒÐŒâ\>>}}

\def\ŒÕŒàŒÓŒá{ŒíŒàŒÓ~Œá$^{-1}$}
\def\ŒÕŒàŒÓŒáŒÜ{ŒíŒàŒÓ~ŒáŒÜ$^{-2}$~Œá$^{-1}$}

\begin{document}

{\footnotesize Astronomy Letters, Vol. 32, No. 1, 2006, pp. 18-28.
Translated from Pis'ma v Astronomicheskii Zhurnal, Vol. 32, No. 1, 2006,
pp. 20-31.  Original Russian Text Copyright \copyright\, 2006 by
T. I. Larchenkova, S. M. Kopeikin}


\title{\bf Shapiro Effect as a Possible Cause of the Low-Frequency Pulsar
Timing Noise in Globular Clusters}

\author{\bf \hspace{-1.3cm}\copyright\, 2006 \ \
   T.I.Larchenkova\affilmark{1}$^{\,*}$ and S.M.Kopeikin\affilmark{2}}

\affil{
$^1$ {\it Astrospace Center, Lebedev Physical Institute, Russian Academy
of Sciences, Profsoyuznaya ul. 84/32, Moscow, 117997 Russia}
$^2$ {\it Department of Physics and Astronomy, University of Missouri
Columbia, Columbia, MO 65211, USA}
}

\vspace{2mm}

\sloppypar
\vspace{2mm}
\noindent


{A prolonged timing of millisecond pulsars has revealed
low-frequency uncorrelated (infrared) noise, presumably of
astrophysical origin, in the pulse arrival time (PAT) residuals for
some of them. Currently available pulsar timing methods allow the
statistical parameters of this noise to be reliably measured by
decomposing the PAT residual function into orthogonal Fourier
harmonics. In most cases, pulsars in globular clusters show a
low-frequency modulation of their rotational phase and spin rate.
The relativistic time delay of the pulsar signal in the curved space
time of randomly distributed and moving globular cluster stars (the
Shapiro effect) is suggested as a possible cause of this modulation.
Extremely important (from an astrophysical point of view)
information about the structure of the globular cluster core, which
is inaccessible to study by other observational methods, could be
obtained by analyzing the spectral parameters of the low-frequency
noise caused by the Shapiro effect and attributable to the random
passages of stars near the line of sight to the pulsar. Given the
smallness of the aberration corrections that arise from the
nonstationarity of the gravitational field of the randomly
distributed ensemble of stars under consideration, a formula is
derived for the Shapiro effect for a pulsar in a globular cluster.
The derived formula is used to calculate the autocorrelation
function of the low-frequency pulsar noise, the slope of its power
spectrum, and the behavior of the $\sigma_z$ statistic that
characterizes the spectral properties of this noise in the form of a
time function. The Shapiro effect under discussion is shown to
manifest itself for large impact parameters as a low-frequency noise
of the pulsar spin rate with a spectral index of $n=-1.8$ that
depends weakly on the specific model distribution of stars in the
globular cluster. For small impact parameters, the spectral index of
the noise is $n=-1.5$. \copyright\, 2006 Pleiades Publishing Inc.}


\vfill

{$^{*}$ E-mail: tanya@lukash.asc.rssi.ru}
\newpage
\thispagestyle{empty}
\setcounter{page}{1}

\section*{INTRODUCTION}

Explaining the astrophysical origin of the low frequency
uncorrelated pulsar timing noise is an extremely important, but
challenging problem of modern pulsar astronomy. It is well known
that millisecond and binary pulsars can be used as the most stable
natural frequency standards (Davis et al. 1985; Rawley et al. 1987;
Kaspi et al. 1994; Petit and Tavella 1996; Ilyasov et al. 1998;
Kopeikin 1999). However, obtaining an unbiased and statistically
significant estimate of their stability depends considerably on the
accurate determination of the parameters of the random low-frequency
noise that is detected in the pulse arrival time (PAT) residuals
observed on long time scales. In most cases, the low-frequency
(infrared) noise can be described by an additive set of components
each of which is described by a simple model of the autocovariance
function with a power-law spectrum, $S(f) = A_n f^{-n}$, where the
index $n$ takes on integer values from unity and higher (Kopeikin
1997). Noise of different physical natures corresponds to different
spectral indices. For this reason, studying the spectral parameters
of the infrared noise is an important problem of observational
astrophysics. It enriches significantly the metrological studies of
the pulsar time scale and encompasses many divisions of modern
astrophysics, the most important of which are cosmology and
gravitational wave astronomy (Kopeikin 1997a; Lorimer 2001).

A self-consistent approach to estimating the stability of the
rotational and orbital parameters of pulsars and pulsar noise models
has been developed by many authors (Groth 1975; Cordes 1980;
Bertotti et al. 1983; Blandford et al. 1984; Taylor 1991; Kopeikin
1997a, 1997b, 1999; Ilyasov et al. 1998; Kopeikin and Potapov 2004).
They solved the main theoretical questions of the analytical
observational data processing technique in the time domain. The data
processing in the time domain is more informative than that in the
frequency domain, since both the stationary and nonstationary noise
components are taken into account in the former case, while an
adequate data analysis in the frequency domain and obtaining
unbiased estimates of the measured parameters of the time series are
possible only for a stationary random process.

It should be noted that the accuracy of determining the pulsar PATs is
currently very high and approaches $\sim50$ ns on a time scale of several
years (Bailes 2003). Therefore, we need a high-precision model for
processing the PATs that are affected by many factors, such as the Earth's
orbital and rotational motions, the proper motion of the pulsar, the
gravitational potentials of the Solar system and other gravitating objects
at the point of observation and along the pulse propagation path, and the
spatial distribution of the interplanetary and interstellar media. The
classical and relativistic effects related to the orbital motion of the
pulsar around the barycenter of the binary and the proper motion of the
latter in space, the propagation of the radio emission from the pulsar in
the gravitational field and atmosphere of its companion, and the precession
of the pulsar spin axis, which causes the pulsar pulse shape to change with
time, are added for pulsars in binary systems.

Based on the relativistic theory of astronomical reference frames and time
scales (Kopeikin 1988, 1989a, 1989b; Brumberg and Kopeikin 1989, 1990),
whose improved and extended version was taken by the General Assembly of
IAU-2000 as the basis for relativistic astronomical algorithms and was
described in detail in review papers (Soffel et al. 2003; Kopeikin and Vlasov
2004), Doroshenko and Kopeikin (1990, 1995) developed a pulsar timing
algorithm that includes the above effects, to within 10 ns, and that is
suitable for processing the observations of both single and binary pulsars.
This theory was effectively used to systematically develop the TIMAPR
software package (Doroshenko 1997), which is designed for pulsar data
processing (Larchenkova and Doroshenko 1995).  The independent TEMPO code
was developed at the Princeton University, USA (Taylor and Weisberg
1989). Both codes are widely used by Russian and foreign scientists at
various radio observatories worldwide (Lorimer 2001).

The standard procedure for estimating the pulsar parameters on fairly short
time scales is based on the assumption that white noise dominates in the PAT
residuals. However, a fairly long monitoring reveals components of the
correlated infrared noise of astrophysical origin almost in all pulsars,
whose spectrum differs from the white noise spectrum (Cordes and Downs 1985;
D'Alessandro et al. 2001). The spectrum of this correlated (infrared) noise
diverges at zero frequency. This infrared catastrophe compels the researches
to reconsider the standard methods of spectral analysis and forces them to
resort to various kinds of regularization procedures that allow the
divergence to be avoided when modeling the noise spectrum (Kopeikin and
Potapov 2004).

One of the most important problems in modern pulsar timing is to separate
out the infrared noise in the PAT residual spectra and to determine its
amplitude and spectral index. Although this is a challenging problem, its
solution can provide substantial information about the physical processes
inside neutron stars and in the interstellar medium in the path of pulsar
pulse propagation and help to detect low-frequency gravitational waves and
other, no less interesting gravitational effects. The most suitable objects
for solving this problem are millisecond pulsars with a very high spin
stability and, as a result, a low level of intrinsic rotational noise
(Guinot and Petit 1991). Millisecond pulsars were discovered in various
regions of our Galaxy.  However, the population of millisecond pulsars in
globular star clusters is currently most representative.  The first
millisecond pulsar was discovered in 1987 in the core of the globular
cluster M28 (Lyne et al. 1987). A systematic search for pulsars both in M28
and in other globular star clusters began after this discovery. Twenty four
globular clusters in which more than 80 pulsars have been discovered are
known to date (Freire 2004). The globular clusters 47 Tuc (NGC 104) and M15
(NGC 7078), which contain, respectively, 22 and 8 pulsars (Taylor et
al. 1993; Camilo et al. 2000; Freire 2004), and the globular cluster Terzan
5, in which 26 pulsars were discovered (Ransom et al. 2005), are the
record-holders in the number of discovered pulsars.

If the inherent causes of the pulsar spin instability are ignored,
then three physical causes of the low frequency timing noise
produced by external effects can be suggested. First, the stochastic
Shapiro effect, i.e., the random spread in the total time it takes
for the pulsar radio signal to pass through the fluctuating
gravitational field of the globular cluster stars that are
distributed and move randomly in a certain vicinity of the line of
sight to the pulsar; second, the stochastic gravitational
perturbations in the pulsar velocity and acceleration produced by
close passages of globular cluster stars near the pulsar itself
(Joshi and Rasio 1997; Rodin 2000); and, third, the random
fluctuations of the interstellar medium (Smirnova and Shishov 2001).
In this paper, we concentrate on detailed calculations and analysis
of the spectrum of the pulsar noise produced by the stochastic
Shapiro effect. The amplitude of this noise depends significantly on
the radial star density distribution inside the cluster, which is
unknown and model-dependent in most cases. From general
considerations, we may expect the noise amplitude to be comparable
to the gravitational radius of a typical cluster star, i.e.,
$\sim10$ $\mu$s. However, this value may be an order of magnitude
higher, since the noise from the Shapiro effect has a cumulative
property and is directly proportional to the number of stars on the
line of sight to the pulsar that produce this noise. There is no
doubt that the increase in the noise amplitude correlates with the
characteristic time scale of its manifestation.

The random process produced by star passages should be considered as a
special case of gravitational lensing. When applied to microarcsecond
astrometry, this case was considered by Sazhin et al. (1998, 2001). In this
paper, we calculate the noise spectrum, as applied to pulsar
observations. Knowledge of the theoretical power spectrum of the stochastic
Shapiro effect for a pulsar in a globular cluster will allow us, in the case
of its direct measurement, to obtain very important astrophysical
information about the structure of the globular cluster core, which is
inaccessible to other observational methods, and to analytically extend the
Salpeter mass function for globular clusters toward the low-mass stars
comparable in mass to Jupiter. Observation of the stochastic Shapiro effect
will also allow the mass of the dark matter that is possibly concentrated
near the globular cluster cores to be estimated.

Kopeikin and Schafer (1999) were the first to derive an exact formula for
the Shapiro effect produced by a moving massive body and showed that the
positions of the light-deflecting gravitating bodies (stars) should be taken
not at the pulsar pulse arrival time to the observer, but at the
corresponding delayed time due to the finite speed of propagation of the
gravitational perturbation (Kopeikin 2001). This relativistic effect has
received convincing experimental confirmation through high-precision VLBI
measurements of the deflection of light from a quasar by the gravitational
field of a moving Jupiter (Famalont and Kopeikin 2003). We took into account
this effect when deriving an expression for the total relativistic delay
time of the pulsar signal in the gravitational field of arbitrarily moving
globular cluster stars.

This paper is structured as follows. In the next section, we briefly
consider the pulsar timing model and the procedure for estimating its
parameters in the presence of low-frequency noise and introduce the concept
of $\sigma_z$ statistic, which serves as a quantitative measure of pulsar
instability. Subsequently, we write out a formula for the cumulative Shapiro
effect for a pulsar in a globular cluster and an expression for the
$\sigma_z$ statistic in the case where the physically significant noise is
attributable to the Shapiro effect from random passages of stars near the
line of sight. In conclusion, we use the $\sigma_z$ statistic to estimate
the noise spectrum and discuss prospects for numerically analyzing this
effect.

\section*{PULSAR TIMING, ESTIMATION OF PULSAR PARAMETERS, AND THE $\sigma_z$
STATISTIC}

Let the observations begin at time $t_0$. The rotational phase of a pulsar
is specified by a time polynomial:
\begin{equation}
N(t) = \nu_p\Im + \frac{1}{2}\dot\nu_p\Im^2 +
\frac{1}{6}\ddot\nu_p\Im^3 + \frac{1}{24}\dddot\nu_p\Im^4 + \nu_p
\phi_2(\Im) + O(\Im^5)
\end{equation}

where $\Im=\Im(t)$ is the pulse emission time in the proper time
scale of the pulsar; $t$ is the pulse arrival time in the
barycentric time scale of the Solar system; $\nu_p$, $\dot\nu_p$,
$\ddot\nu_p$ and $\dddot\nu_p$ are the pulsar spin rate and its
derivatives taken at time $\Im=0$; and $\phi_2(\Im)$ is the
intrinsic noise of the pulsar rotational phase, spin rate, and its
time derivatives.

Note that the pulsar proper time $\Im$ is not directly observable.
The barycentric time, $t$, which is related to the proper time of
the observer (atomic time) by a relativistic transformation
(Kopeikin 1989b; Brumberg and Kopeikin 1990; Kopeikin and Vlasov
2004), is a measurable quantity. The relationship between the pulsar
proper time $\Im$ and the barycentric time $t$ is established by
solving the equations for the light geodesics that describe the
propagation of the radio pulse from the pulsar to the observer
(Kopeikin 1990; Doroshenko and Kopeikin 1990). This relationship
allows the observed pulsar rotational phase $N(t)$ to be expressed
as a function of the barycentric time $t$ (Doroshenko and Kopeikin
1990; Kopeikin 1999):

\begin{equation}
N(t) = N_0 + \nu t + \frac{1}{2}\dot\nu t^2 + \frac{1}{6}\ddot\nu t^3
+\nu\varepsilon(t),
\end{equation}

where $N_0$ is the initial rotational phase of the pulsar; $\nu$, $\dot\nu$
and $\ddot\nu$ are the barycentric pulsar spin rate and its derivatives at
the initial epoch of observations $T$; and $\varepsilon(t)$ is the total,
physically significant additive noise of the pulsar rotational phase, with
the intrinsic noise of the pulsar $\phi_2(t)$ being one of its components.
We emphasize that the noise $\varepsilon(t)$ is produced by both external
and internal factors. By the external factors we mean all those factors that
affect the propagation of the pulsar radio pulse in a random way. These also
include the noise from the passages of globular cluster stars, which will be
considered in detail in the next sections. The internal factors responsible
for the pulsar noise are related to the pulsar spin mechanism and are not
considered here. Lyne and Graham-Smith (2004) gave a comprehensive overview
of the causes of the pulsar spin instability.

The PAT residual $r(t)$ is the difference between the observed pulsar
rotational phase, $N_{obs}$, and its theoretical value, $N(t,\theta)$,
predicted by the timing model and specified by Eq. (2) divided by the pulsar
rotation rate, $\nu$:
$$ r(t,\theta)=\frac{N_{obs}-N(t,\theta)}{\nu}, $$
where $\theta=\{\theta_a, a=1,2,...,. . .,k\}$ denotes a set $k$ of
measurable parameters ($k=4$ in the model represented by Eq. (2)).

Let us introduce the quantities $\beta_a \equiv
\theta^{*}_{a}-\hat{\theta}_a$ that are the corrections to the
unknown true values of the parameters  $\hat{\theta}_a$ and the
timing model fits
$$\psi_a (t) = \left[\frac{\partial N}{\partial
\theta_a}\right]_{\theta=\theta^*}, $$ where $\theta^{*}_{a}$ are
the least-squares estimates of the parameters (Kopeikin 1999). The
parameters and fits are given in the table.

Assuming that $m$ equally spaced and comparable (in accuracy) pulse
arrival times are measured for $N$ complete rotations of the pulsar
around its axis, we have $mN$ residuals: $r_i\equiv r(t_i)$, where
$i=1,2,...,mN$. The least-squares method yields the parameters
$\beta_a(\tau)$ (Bard 1974):
$$\beta_a(\tau) = \sum_{b=1}^{4} \sum_{i=1}^{mN} L^{-1}_{ab}
\psi_b(t_i) \varepsilon(t_i), ~~~~a=1,2,3,4,$$

where the information matrix
$$L_{ab}(\tau) = \sum_{i=1}^{mN} \psi_a (t_i) \psi_b (t_i),$$ $\tau$
is the total observing time. The correlation matrix of the
parameters is defined as $M_{ab}\equiv\langle\beta_a\beta_b\rangle$,
where the angular brackets denote an ensemble average.

The behavior of the PAT residuals can be described best using the
so-called $\sigma_z$ statistic. It is defined as the weighted
root-mean-square value of the coefficients of the cubic terms in the
time polynomial fitted to the observed pulsar phase divided into the
segments that correspond to equal time intervals $\tau$ (Matsakis et
al. 1997). The $\sigma_z$ statistic is formally defined (Matsakis et
al. 1997) as
$$\sigma_z(\tau)=\frac{\tau^2}{2\sqrt{5}}\sqrt{M_{44}},$$
where $M_{44}$ is a diagonal element of the correlation matrix
$M_{ab}$. This diagonal element is defined by

\begin{equation}
M_{44}(\tau) = \sum_{c=1}^{4}\sum_{d=1}^{4} L^{-1}_{4c} L^{-1}_{4d}
\times \left[\sum_{i=1}^{mN}\sum_{j=1}^{mN}
\psi_c(t_i)\psi_d(t_i)\Re(t_i,t_j)\right]
\end{equation}

where $\Re(t_i,t_j)=\langle\varepsilon(t_i)\varepsilon(t_j)\rangle$
is the autocorrelation function of the random process
$\varepsilon(t)$. For a stationary noise process, the
autocorrelation function depends not on the individual times $t_i$
and $t_j$ , but only on their difference $\tau$:
$$\Re(t,\tau)=\langle\varepsilon(t+\tau)\varepsilon(t)\rangle\equiv\Re(\tau),$$
where $t=t_j$, $\tau=t_i-t_j$.

\vspace{5mm}

\begin{center}
List of basic functions and parameters used to fit the
parameters in the pulsar timing model specified by Eq.(2)

\begin{tabular}{c|c}
\hline
~~~~~~~~~~~~~Parameter~~~~~~~~~~~~~ & ~~~~~~~~~~~~~~Fit~~~~~~~~~~~~~ \\
\hline
$\beta_1=\delta N_0/\nu$      & $\psi_1(t)=1$ \\
$\beta_2=\delta\nu/\nu$      & $\psi_2(t)=t$ \\
$\beta_3=\delta\dot\nu/\nu$  & $\psi_3(t)=t^2$ \\
$\beta_4=\delta\ddot\nu/\nu$ & $\psi_4(t)=t^3$ \\
\hline
\end{tabular}
\end{center}

\section*{THE SHAPIRO EFFECT FOR A PULSAR IN A GLOBULAR CLUSTER}

The relativistic time delay of an electromagnetic signal in a
static, spherically symmetric gravitational field of a point mass is
called the Shapiro effect (Shapiro 1964). A generalization of the
formula for the Shapiro effect for the propagation of light in a
variable gravitational field of an arbitrarily moving body was found
by Kopeikin and Schafer (1999) and experimentally confirmed by
Fomalont and Kopeikin (2003). In this paper, we use the
Kopeikin-Schafer formula.

\subsection*{Formula for the Generalized Shapiro Effect}

The expression for the travel time of an electromagnetic signal
between fixed points, ($t_0, \vec{x}_0$) and ($t, \vec{x}$), in the
gravitational field of an arbitrarily moving body (star) derived by
Kopeikin and Schafer (1999) is
$$t-t_0 = |\vec{x}-\vec{x}_0|+\Delta(t, t_0),$$
where $|\vec{x}-\vec{x}_0|$ is the coordinate distance in the
background Euclidean space between the two points, $\vec{x}_0$ and
$\vec{x}$; and $\Delta(t, t_0)$ is the relativistic time delay
attributable to the gravitational field of the moving bodies.

For a pulsar in a globular cluster, the Shapiro delay produced by
the gravitational field of the moving cluster stars can be written
as (Kopeikin and Schafer 1999)
\begin{equation}
\Delta(t, t_0) = - \sum_{a=1}^{N} \frac{2GM_a}{c^3} \ln
\frac{r_a-(\vec{k}_0\vec{r}_a)}{r_{0a}-(\vec{k}_0\vec{r}_{0a})},
\end{equation}
$$\vec{r}_a = \vec{x}(t)-\vec{x}_a(s),$$
$$\vec{r}_{0a} = \vec{x}_0(t_0)-\vec{x}_a(s_0),$$
where we discarded the terms that are proportional to the velocity
of the bodies and that appear in the amplitude of the logarithmic
function, because they are small. The coordinates of the gravitating
bodies are calculated at the delayed times $s$ and $s_0$, which are
defined by the equations for isotropic gravitational field
characteristics:
\begin{eqnarray}
t = s + |\vec{x}(t)-\vec{x}_a(s)|,\\
t_0 = s_0 + |\vec{x}_0(t_0)-\vec{x}_a(s_0)|. \nonumber
\end{eqnarray}

Here, $t_0$ is the photon emission time, $\vec{x}$ specifies the
observer's position relative to the globular cluster barycenter
(GCB) at the observation time of the pulsar radio pulse, $\vec{x}_a$
specifies the position of star a relative to the GCB, $\vec{x}_0$
specifies the pulsar position relative to the GCB at the radio pulse
emission time, and the unit vector $\vec{k}_0$ specifies the
direction of the rectilinear, gravitationally unperturbed radio
pulse motion from the pulsar to the observer and is defined by
$$\vec{k}_0=\frac{\vec{x}-\vec{x}_0(t_0)}{|\vec{x}-\vec{x}_0(t_0)|}.$$

Let us expand $r_a$, $\vec{r}_a$, and $\vec{k}_0$ in a Taylor power
series of $x_a/R \ll 1$, where $R = |\vec{x}(t)|$ is the distance
from the observer to the GCB, much larger than the cluster size:
$$\vec{k}_0 = \vec{K}-\vec{K}\times(\vec{\xi}\times\vec{K})+O(\vec{\xi}^2),$$
$$\vec{r}_a = R-R(\vec{K}\vec{\xi}_a)+\frac{1}{2}R(\vec{\xi}_a\times\vec{K})^2+O(\vec{\xi}^{3}_{a}),$$
where $\vec{K} = \vec{x}(t)/|\vec{x}(t)|$ is a unit vector,
$\vec{\xi} = \frac{\vec{x}_0(t_0)}{|\vec{x}(t)|}$, and $\vec{\xi}_a
= \frac{\vec{x}_a(s)}{|\vec{x}(t)|}$.

Let $T$ be a fixed start time of observations. Let us expand
$\vec{x}(t)$ and $\vec{x}_a(s)$ in a Taylor power series of $(t-T)$,
provided that the following conditions are satisfied:
$|\vec{x}(T)|\gg |\vec{v}(T)(t-T)|\gg|\dot{\vec{v}}(T)(t-T)^2|$.
Retaining only the linear (in velocities) terms and using the
following notation: $R_0=|\vec{x}(T)|$, $\vec{K}_0=\vec{R}_0/R_0$ is
a unit vector, $\vec{v}$ is the observer's velocity, $\vec{v}_a$ is
the velocity of star a, and $\vec{v}_0$ is the pulsar velocity, we
then obtain
\begin{eqnarray}
\vec{x}(t)=\vec{x}(T)+\vec{v}(T)(t-T)+\ldots, \nonumber \\
\vec{x}_a(s)=\vec{x}_a(S)+\vec{v}_a(s-S)+\ldots, \nonumber \\
R=R_0+(\vec{K_0}\vec{v}(T))(t-T)+\ldots, \nonumber \\
\vec{K}=\vec{K}_0+\left[\frac{\vec{v}(T)-\vec{K}_0(\vec{K}_0\vec{v}(T))}
{R_0}\right](t-T)+\ldots, \nonumber
\end{eqnarray}
where $S$ is the delayed time related to the start time of
observations $T$ by the gravitational cone equations (5).

Using the equations for gravitational field characteristics (5), we
obtain the following expression for the numerator of the fraction in
the argument of the logarithm in Eq. (4) after algebraic
transformations:
$$\vec{r}_a-(\vec{k}\vec{r}_a)=1+2\frac{\vec{d}_a\vec{v}_a(S)}{\vec{d}_a\vec{x}_a(S)}(t-T)+\ldots,$$
where
$\vec{d}_a=(\vec{K}_0\times\frac{\vec{x}_a}{R_0})\times\vec{K}_0$ is
the impact parameter of the radio pulse trajectory with respect to
the GCB.

Acting in a similar way, we expand $\vec{r}_{0a}$ and $r_{0a}$ in a
Taylor series, retaining only the linear (in velocity) terms, and
obtain an expression for the denominator of the fraction in the
argument of the logarithm in Eq. (4) after simple transformations.
Finally, once all terms have been reduced to the same time $T_0$,
which defines the pulsar pulse emission time corresponding to the
start of observations $T$, we obtain the ultimate expression for the
random process produced by the relativistic delay during the
propagation of a radio pulse in a variable gravitational field of
moving globular cluster stars:
\begin{eqnarray}
\tilde{\varepsilon}(t)\equiv\Delta(t,t_0)-\Delta(T,T_0) \\
=\sum_{a=1}^{N}\frac{2GM_a}{c^3} \left[\ln \left\{ 1+2(t-T) \frac
{\vec{d}_a(T_0)\vec{v}_a(T_0)}
{\vec{d}_a(T_0)\vec{x}_a(T_0)}\right\} -\ln\left\{1+
\frac{(-\vec{q}-\vec{K}_0)\vec{V}_{0a}(T_0)}{Q_{0a}+\vec{K}_0\vec{Q}_{0a}}
(t-T)\right\}\right], \nonumber
\end{eqnarray}
where $\vec{V}_{0a}=\vec{v}_0(T_0)-\vec{v}_a(T_0)$,
$\vec{Q}_{0a}=\vec{x}_a(T_0)-\vec{x}_0(T_0)$,
$Q_{0a}=(\vec{Q}_{0a}\vec{Q}_{0a})^{1/2}$,
$\vec{q}=\vec{Q}_{0a}/Q_{0a}$ is a unit vector.

\subsection*{Autocorrelation Function of the Shapiro Effect}

Let us pass to a new model of the random process $\varepsilon(t)$
defined by the formula
$$\varepsilon(t)=\tilde{\varepsilon}(t)-\langle\tilde{\varepsilon}(t)\rangle,$$
where the angular brackets denote a statistical ensemble average,
and we assume that the mean value is
$\langle\varepsilon(t)\rangle=0$. The autocorrelation function in
Eq. (3) can then be written as
\begin{equation}
\Re(t,\tau)=\int dm_a d\vec{x}_a d\vec{v}_a
f(m_a,\vec{x}_a,\vec{v}_a)  \varepsilon(t,m_a,\vec{x}_a,\vec{v}_a)
\varepsilon(t+\tau,m_a,\vec{x}_a,\vec{v}_a),
\end{equation}
where we assume that the statistical ensemble of stars is defined by
uncorrelated parameters, so the distribution function can be fitted
by the product of three statistically independent distribution
functions:
$$f(m_a,\vec{x}_a,\vec{v}_a)=Af(m_a)f(\vec{x}_a)f(\vec{v}_a);$$
we find the normalization numerical coefficient $A$ from the
condition
$$A=\int dm_a d\vec{x}_a d\vec{v}_a
f(m_a,\vec{x}_a,\vec{v}_a)=1.$$

\noindent No integration limits are specified in Eq. (7), but we
assume that they are known (see below) and specify the range of
statistical ensemble parameters.

A globular cluster is characterized by two quantities: the cluster
core radius $r_c$, which is defined as the distance at which the
surface brightness is half its central value, and the tidal radius
$r_t$, at which the surface brightness is zero. Given the mass of
the Galaxy $M_G$, the cluster mass $M_c$, and the Galactocentric
distance of the cluster center $R_G$, the tidal radius can be
calculated as follows: $r^3_t = \frac{M_c}{2M_G} R^3_G$. Equation
(7) is integrated in the following limits: $m_a =
[0.1M_\odot-10M_\odot]$, $|\vec{v}_a|=[0-4\sigma]$ and $|\vec{x}_a|=
[0-r_t]$, where $\sigma$ is the stellar velocity dispersion of the
cluster, and $M_\odot$ is the solar mass.

We use the Salpeter function $f(m_a)\sim m^{-2.35}_{a}$ as the mass
function of globular cluster stars. Let us consider two model
density distributions of a globular cluster: the model of an
isothermal sphere with a core and the King model. In spherical
coordinates, the density distribution for the model of an isothermal
sphere with a core is (Spitzer 1990)
\begin{equation}
f(r_a) = \frac{\rho_0r^2_c}{r^2_c+r^2_a},
\end{equation}
where $\rho$ is the core density of the globular cluster. The
velocity distribution in the model of an isothermal sphere has a
Gaussian (Maxwellian) profile,
\begin{equation}
f(v_a) = \frac{1}{2\pi\sigma^2} \exp \left[ - \frac
{v^2_{ax}+v^2_{ay} + v^2_{az}} {2\sigma^2}\right].
\end{equation}
It is the Maxwellian distribution that is typical of unrelaxed
systems, which the globular clusters in our Galaxy are. However, the
model of an isothermal sphere is not devoid of shortcomings. The
main shortcoming is the assumption of its infinite radius in
geometric space and velocity space, which is not physically
justified. This shortcoming can be eliminated by "truncating" the
argument of the velocity distribution function (9) by the
characteristic star escape velocity $v_e$. This allows the
isothermal sphere to be made finite in velocity space while
preserving the isotropic Maxwellian distribution. From physical
considerations in the velocity range $v_a>v_e$, the distribution
function $f(v_a)$ must be close to zero and can be fitted by a
truncated Maxwellian function:
$$f(v_a)=\frac{e^{-v^2_a/\sigma^2}-e^{-v^2_e/\sigma^2}}{1-e^{-v^2_e/\sigma^2}},$$
where the normalization was chosen in such a way that $f(v_a)=1$ at
$v_a=0$. The models of star clusters based on the truncated
Maxwellian distribution were calculated by King (1966) and are
called the King models. The density distribution in the King models
is also bounded in space by the tidal radius $r_t$ and is defined as
(King 1966)
\begin{equation}
f(r_a) = \rho_0 \frac {(1+\Gamma^2) \arccos
\sqrt{\frac{1+(r_a/r_c)^2}{1+\Gamma^2}} -
\sqrt{\Gamma^2-(r_a/r_c)^2} \sqrt{1+(r_a/r_c)^2}}
{\left[1+(r_a/r_c)^2\right]^{3/2} \left[(1+\Gamma^2) \arccos
\sqrt{\frac{1}{1+\Gamma^2}}-\Gamma \right]},
\end{equation}
where $\Gamma\equiv r_t/r_c$. For the subsequent calculations, we
use the globular cluster 47 Tucanae, in which the largest number of
pulsars have been discovered to date. This cluster has the following
parameters: $r_c = 0.52$ pc, $\rho_0 = 6\times10^4 M_\odot/{\rm
pc}^3$, $r_t = 60.3$ pc, and $\sigma= 10$ km s$^{-1}$; the distance
to the cluster center is $R_0 = 4.1$ kpc (Harris 1996). To determine
the slope of the power spectrum for the noise process under study
and the behavior of the $\sigma_z$ statistic, we must calculate the
integral in Eq. (7). Analysis indicates that this integral cannot be
calculated analytically in general form and the effect under study
must be simulated numerically, which is done here. The random
process attributable to the Shapiro effect can be estimated
analytically if we restrict our analysis to star passages far from
the line of sight to the pulsar. The random process under study can
be separated into long- and short-period parts. The long-period part
of the process corresponds to the case where the impact parameter da
of the pulsar pulse trajectory is large for the cluster stars
(analytical case). The short-period part of the process corresponds
to the case of small impact parameters.

\subsection*{Large Impact Parameters}

Let us make the simplifying assumption that the distributions of
globular cluster stars in space, velocity, and mass are
uncorrelated. This simplifies significantly the form of the
distribution function in Eq. (7):
\begin{equation}
f(m_a,\vec{x}_a,\vec{v}_a,m_b,\vec{x}_b,\vec{v}_b)=\delta(m_a -
m_b)\delta(\vec{x}_a - \vec{x}_b) \delta(\vec{v}_a-\vec{v}_b)
f(m_a,\vec{x}_a,\vec{v}_a), (a\ne b)
\end{equation} where $\delta(y)$ is
the Dirac delta function. The assumption about large impact
parameters of the pulsar pulse trajectory relative to the globular
cluster stars allows the expression for the random process under
study (6) to be expanded in a Taylor power series of $(t - T)$:
\begin{equation}
\tilde{\varepsilon}(t) = \alpha(t-T) + \beta(t-T)^2 + \gamma (t-T)^3
+\ldots,
\end{equation}
where $T$ is the initial epoch of observations and the coefficients
$\alpha$, $\beta$ and $\gamma$ are defined by
\begin{eqnarray}
\alpha=\sum_{a=1}^{N} \frac {2GM_a}{c^3} \left[ \frac {2\vec{d}_a
\vec{v}_a}{\vec{d}_a \vec{x}_a} - \frac
{(-\vec{q}-\vec{K}_0)\vec{V}_{0a}} {Q_{0a}+\vec{K}_0\vec{Q}_{0a}}
\right], \nonumber \\
\beta=\sum_{a=1}^{N} \frac {2GM_a}{c^3} \frac{1}{2} \times \left\{
\left[ \frac {(-\vec{q}-\vec{K}_0)\vec{V}_{0a}}
{Q_{0a}+\vec{K}_0\vec{Q}_{0a}} \right]^2 - \left(\frac {2\vec{d}_a
\vec{v}_a}{\vec{d}_a \vec{x}_a} \right)^2 \right\}, \nonumber \\
\gamma=\sum_{a=1}^{N} \frac {2GM_a}{c^3} \frac{1}{3} \times
\left\{\left(\frac {2\vec{d}_a \vec{v}_a}{\vec{d}_a \vec{x}_a}
\right)^3 - \left[ \frac {(-\vec{q}-\vec{K}_0)\vec{V}_{0a}}
{Q_{0a}+\vec{K}_0\vec{Q}_{0a}} \right]^3 \right\}. \nonumber
\end{eqnarray}

In what follows, it will suffice to restrict the analysis to only
the first term in expansion (12), which is linear in time $(t - T)$.
Let us write the autocorrelation function for it using the
expression for the distribution function (11):

\begin{eqnarray}
\Re(t_1, t_2) = \int dmd\vec{x}d\vec{v}\alpha^2(m,\vec{x},\vec{v})
\times f(m,\vec{x},\vec{v})(t_1-T)(t_2-T) = \\
\int dmd\vec{x}d\vec{v}\alpha^2(m,\vec{x},\vec{v})
f(m,\vec{x},\vec{v}) \times [t_1 t_2 - T(t_1+t_2)+T^2]. \nonumber
\end{eqnarray}

Let us designate $\tau = |t_2 - t_1|$ and $t_+ = (t_1 + t_2)$. These
designations allow the autocorrelation function (13) to be
represented as

\begin{equation}
\Re(t_1, t_2) = \int dmd\vec{x}d\vec{v} \alpha^2(m,\vec{x},\vec{v})
f(m,\vec{x},\vec{v}) \times \left( t^2_+ - \frac{\tau^2}{4} -2t_+T +
T^2 \right).
\end{equation}

In this formula, the terms proportional to $t^2_+$, $t_+T$, $T^2$
constitute the nonstationary part of the noise. The terms
proportional to $t_+$ contribute only to the initial rotational
phase of the pulsar, while $t^2_+$, $t_+T$, $T^2$ are equal to the
products of the fits. According to the theorem that was proved by
Kopeikin (1999), these products of the fits in the structure of the
autocorrelation function do not contribute to the {\it PAT
residuals, which depend only on the stationary part of the noise}.
Thus, only the stationary component $\frac{\tau^2}{4}$ in Eq. (14)
contributes to the PAT residuals. For this reason, the random
process attributable to the regular part of the Shapiro effect for
pulsars in globular clusters will manifest itself mainly as the
noise of the pulsar spin rate.

\subsection*{Numerical Model for the Pulsar Noise Produced by the
Shapiro Effect}

As we noted above, knowledge of the autocorrelation function (7)
allows us to determine the slope of the power spectrum for the noise
process and the behavior of the $\sigma_z$ statistic. We calculated
the triple integral (7) numerically. We constructed two numerical
models for the density distribution of the globular cluster 47 Tuc
described by Eqs. (8) and (10).We also considered two extreme cases:
small and large impact parameters.

{\bf Large impact parameters.} We calculated the autocorrelation
function in two ways: (1) in a spherical coordinate system with the
origin at the GCB and (2) in a Cartesian coordinate system. We
equated the volumes of the integration space in both coordinate
systems and took into account the analytic singularities of the
integrand by choosing the appropriate path of integration. Indeed,
it follows from the form of the expression for the Shapiro delay (4)
attributable to the gravitational field of the moving cluster stars
that a star on the line of sight produces an in.nite signal delay.
For a numerical model, this necessitates cutting out a cylinder of
minimum radius $r_{cl}$ in the direction of the unit vector
$\vec{k}_0$. Given the radius of the cutout cylinder, we can easily
estimate the cluster mass, which is excluded from the analysis. For
example, for $r_{cl}\sim 1$AU, it is $\sim10^{-4}M_\odot$.

The relative accuracy of calculating the integral was determined by
the standard method and is 0.9\% for the King model and 1\% an
isothermal sphere with a core. The autocorrelation function for the
King model and the model of an isothermal sphere with a pulsar at
the cluster center is plotted in Figs. 1 and 2, respectively.

Analysis of the plots of the autocorrelation function indicates that
it is well fitted by a quadratic polynomial for both the King model
and the model of an isothermal sphere. For example, $\Re = 7 \times
10^{-23} + 1.16\times10^{-20}t + 7\times10^{-23}t^2$ for the King
model ($\tau = 3$ yr) and $\Re = 5\times10^{-24}+ 6.12\times
10^{-22}t + 4\times10^{-24}t^2$ for the model of an isothermal
sphere with a core ($\tau= 3$ yr). As would be expected, our
numerical result matches the analytic prediction made in the
previous subsection and confirms the validity of the numerical
integration model.

A stochastic process can be described both in the time domain by a
time-dependent quantity $h(t)$ and in the frequency domain by the
amplitude of the process $H(f)$ as a function of the Fourier
frequency $f$. These two quantities are related by the Fourier
transform:
\begin{eqnarray}
H(f) = \int_{-\infty}^{\infty} H(f)e^{2\pi ift}dt \nonumber \\
h(f) = \frac{1}{2\pi}\int_{-\infty}^{\infty} H(f)e^{-2\pi ift}dt \nonumber \\
\end{eqnarray}

According to the Parseval theorem, the total power of the signal
will be the same when calculated in both the time and frequency
domains:
$$P\equiv \int_{-\infty}^{\infty} |h(t)|^2 dt =
\int_{-\infty}^{\infty} |H(f)|^2 df.$$

If the function $h(t)$ is real, then we can define the onesided
spectral power density as
$$P(f) = 2|H(f)|^2.$$

The total power will then be calculated as an integral of $P(f)$ in
the Fourier frequency limits from 0 to $\infty$.

To analyze the noise process, we numerically determined the spectral
power density $P(f)$ and the spectral slope $n$ using the fast
Fourier transform, whose algorithm has been well developed (Elliott
and Rao 1982). Our calculations show that the slopes of the power
spectrum are $n = -(1.78\pm0.04)$ for the King model and $n = -(1.76
\pm0.04)$ for the model of an isothermal sphere, which are equal
within the computational error limits. In Fig. 3, $\log P$ is
plotted against log f for the King model.

{\bf Small impact parameters.} The random process attributable to
star passages with small impact parameters should be considered as a
special case of gravitational lensing. In our case, the
characteristic impact parameter is equal to the Einstein-Chwolson
radius. For a pulsar at the center of a globular star cluster, the
Einstein-Chwolson radius is de.ned by (Zakharov and Sazhin 1998)
$$R_E = \sqrt{\frac{4 G M_a D_{ds} D_d} {c^2D_s}},$$
where $D_{ds}$ is the distance from the pulsar to the gravitating
body (star), $D_d$ is the distance from the observer to the
gravitating body, and $D_s$ is the distance from the observer to the
pulsar. For the globular cluster 47 Tuc and a gravitating mass of
the order of the solar mass $M_\odot$, the characteristic
Einstein-Chwolson radius is $\sim1$ AU. Therefore, in the direction
of the pulsar specified by the unit vector $\vec{k}_0$, we cut out a
cylindrical volume with a polar radius equal to the
Einstein-Chwolson radius. We calculated integral (7) in a
cylindrical coordinate system with a coordinate grid whose density
increased with decreasing polar radius, since the stars with the
smallest impact parameters make a larger contribution to the noise
process.

The autocorrelation function of the noise process produced by the
Shapiro effect is plotted in Fig. 4 for impact parameters smaller
than 1 AU in the model of an isothermal sphere for three observing
times $\tau$: 1, 3, and 5 yr. The derived time dependences of the
autocorrelation function cannot be fitted by a quadratic time
polynomial, as in the case of large impact parameters, because the
gravitational field acts on the radio pulse produced by close star
passages near the line of sight for a short time. The
autocorrelation function in Fig. 4 is nearly logarithmic and, in all
probability, can be interpreted as a flicker noise (Kopeikin 1997b,
1999). The spectral index of the pulsar noise for small impact
parameters is $n = -(1.55 \pm0.03)$. Figure 5 shows the slope of the
power spectrum for the model of an isothermal sphere for impact
parameters smaller than 1 AU and an observing time of $\tau= 5$ yr.

{\bf Dependence of the noise on cluster core radius.} In conclusion,
let us analyze the dependence of the behavior of the autocorrelation
function for the stochastic Shapiro effect on the globular cluster
core radius. The numerical model described in the previous section
was constructed for the globular cluster 47 Tuc whose core radius is
$r_c = 0.52$ pc. We modeled the behavior of the autocorrelation
function for three different cluster core radii, $r_c = 0.08$, 0.1,
and 0.52 pc. The results of our calculations are shown in Fig. 6. It
is easy to see that in all three cases, the shape of the time
dependence of the autocorrelation function is the same, and, hence,
the slope of the power spectrum does not depend on the cluster core
radius used in our calculations.

\section*{CONCLUSIONS}

The rotational phase modulation of a pulsar in a globular star
cluster may be attributable to the fluctuations in the relativistic
time delay of the pulsar signal caused by the gravitational field
variations due to the motion of the cluster stars. The effect of the
low-frequency timing noise produced by this process can be studied
by accurately determining the behavior of the PAT residuals in the
time and/or frequency domain. Here, we derived an analytical
expression for the Shapiro delay of the radio pulse from a pulsar in
a globular cluster by applying small aberration corrections, of the
order of $(v/c)$, where $v$ is the characteristic velocity of the
stars in the cluster and $c$ is the speed of light.

Assuming that the interactions between the cluster stars are
uncorrelated for large impact parameters, the random process
produced by the stochastic Shapiro effect manifests itself mainly as
the noise of the pulsar spin rate. For small impact parameters, we
numerically simulated the stochastic Shapiro effect by assuming that
the mass distribution of the cluster stars was specified by the
Salpeter function, the velocity distribution of the cluster stars
was Maxwellian, and the globular cluster density was described
either by the model of an isothermal sphere or by the King model.

Our numerical analysis showed that the autocorrelation function of
the noise process produced by the Shapiro effect attributable to the
gravitational field of the moving cluster stars could be fitted by a
quadratic time polynomial for both the model of an isothermal sphere
and the King model at large impact parameters. The slope of the
power spectrum is $n\approx -1.8$ and depends weakly on the model
density distribution of the globular star cluster. For small impact
parameters, the functional time dependence of the autocorrelation
function for the stochastic Shapiro effect is nearly logarithmic and
the slope of the power spectrum decreases and is $n\approx -1.5$.

\section*{ACKNOWLEDGMENTS}

We are grateful to the referees for a careful reading of the paper
and for the suggestions to improve it. T.I. Larchenkova wishes to
thank N.A. Arkhipova, V.N. Lukash, and E.V. Mikheeva for helpful
discussions. The work by T.I. Larchenkova was supported in part by
the Russian Foundation for Basic Research (project no. 04-02-1744).

\section*{REFERENCES}
1. M. Bailes, Astron. Soc. Pac. Conf. Proc. {\bf 302}, 57 (2003).

2. Y. Bard, Nonlinear Parameter Estimation (Academic, New York,
1974; Finansy i Statistika,Moscow, 1979). ).

3. B. Bertotti, B. J. Carr, and M. J. Rees, Mon. Not. R. Astron.
Soc. {\bf 203}, 945 (1983).

4. R. Blandford, R. Narayan, R. W. Romani, Astron. Astrophys. {\bf
5}, 369 (1984).

5. V. A. Brumberg and S. M. Kopeikin, in Reference Frames in
Astromomy and Geophysics, Ed. by J. Kovalevsky, I. I. Meuler, B. G.
Kolaczek, et al. (Kluwer, Dordrecht, 1989), p. 115.

6. V. A. Brumberg and S. M. Kopeikin, Celest. Mech. Dyn. Astron.
{\bf 48}, 23 (1990).

7. F. Camilo, D.R. Lorimer, P. Freire, et al., Astron. Soc. Pac.
Conf. Ser. {\bf 202}, 3 (2000).

8. J. M. Cordes, Astrophys. J. {\bf 237}, 216 (1980).

9. J. M. Cordes and G. S. Downs, Astrophys. J., Suppl. Ser. {\bf
59}, 343 (1985).

10. F.D'Alessandro, A. A. Deshpande, and P.M.McCulloch, Astron.
Astrophys. {\bf 18}, 5 (1997).

11. M. M. Davis, J. H. Taylor, J. M. Weinberg, and D. C. Backer,
Nature {\bf 315}, 547 (1985).

12. O. V. Doroshenko,
http://www.mpifrbonn.mpg.de/div/pulsar/former/olegd/soft.html
(1997).

13. O. V. Doroshenko and S. M. Kopeikin, Astron. Zh. {\bf 67}, 986
(1990) [Sov. Astron. {\bf 34}, 496 (1990)].

14. O. V. Doroshenko and S. M. Kopeikin, Mon. Not. R. Astron. Soc.
{\bf 274}, 1029 (1995).

15. D. F. Elliott and K. R. Rao, Fast Transforms: Algorithms,
Analyses, Applications (Academic, New York, 1982).

16. E. B. Fomalont and S. M. Kopeikin, Astrophys. J. {\bf 598}, 704
(2003).

17. P. C. Freire, http://www.naic.edu/pfreire/GCpsr.html (2004).

18. E. J. Groth, Astrophys. J., Suppl. Ser. {\bf 29}, 443 (1975).

19. B. Guinot and G. Petit, Astron. Astrophys. {\bf 248}, 292
(1991).

20. W. E. Harris, Astron. J. {\bf 112}, 1487 (1996).

21. Yu. P. Ilyasov, S. M. Kopeikin, and A. E. Rodin, Pis'ma Astron.
Zh. {\bf 24}, 228 (1998) [Astron. Lett. {\bf 24}, 275 (1998)].

22. K. J. Joshi and F. A. Rasio, Astrophys. J. {\bf 479}, 948
(1997).

23. V. M. Kaspi, J. H. Taylor, and M. F. Ryba, Astrophys. J. {\bf
428}, 713 (1994).

24. I. R. King, Astron. J. {\bf 71}, 64 (1966).

25. S. M. Kopeikin, Celest. Mech. {\bf 44}, 87 (1988).

26. S. M. Kopeikin, Astron. Zh. {\bf 66}, 1069 (1989à) [Sov. Astron.
{\bf 33}, 550 (1989à)].

27. S. M. Kopeikin, Astron. Zh. {\bf 66}, 1289 (1989b) [Sov. Astron.
{\bf 33}, 665 (1989b)].

28. S. M. Kopeikin, Astron. Zh. {\bf 67}, 10 (1990) [Sov. Astron.
{\bf 34}, 5 (1990)].

29. S. M. Kopeikin, Phys. Rev. D {\bf 56}, 4455 (1997à).

30. S. M. Kopeikin, Mon. Not. R. Astron. Soc. {\bf 288}, 129
(1997b).

31. S. M. Kopeikin, Mon. Not. R. Astron. Soc. {\bf 305}, 563 (1999).

32. S. M. Kopeikin, Astrophys. J. {\bf 556}, L1 (2001).

33. S. M. Kopeikin and I. Y. Vlasov, Phys. Rep. {\bf 400}, 209
(2004).

34. S. M. Kopeikin and V. A. Potapov, Mon. Not. R. Astron. Soc. {\bf
355}, 395 (2004).

35. S. M. Kopeikin and G. Schafer, Phys. Rev. D {\bf 60}, 124002
(1999).

36. T. I. Larchenkova and O. V. Doroshenko, Astron. Astrophys. {\bf
297}, 607 (1995).

37. D. Lorimer, http://www.limingreviews.org/lrr-2001-5.

38. A. G. Lyne and F. Graham-Smith, Pulsar Astronomy (Cambridge
Univ. Press, Cambridge, 2005).

39. A.G. Lyne,A. Brinklow, J.Middleditch, et al., Nature {\bf 328},
399 (1987).

40. D. N. Matsakis, J. H. Taylor, and T. Marshall Eubanks, Astron.
Astrophys. {\bf 326}, 924 (1997).

41. G. Petit and P. Tavella, Astron. Astrophys. {\bf 308}, 290
(1996).

42. S. M. Ransom, J. W. T. Hessels, I. H. Stairs, et al.,
astro-ph/0501230 (2005).

43. L. A. Rawley, J.H. Taylor, M. M. Davis, and D.W. Allan, Science
{\bf 238}, 761 (1987).

44. A. E. Rodin, Candidate Dissertation (Lebedev Phys. Inst.,Moscow,
2000).

45. M. V. Sazhin, V. E. Zharov, and T. A. Kalinina, Mon. Not. R.
Astron. Soc. {\bf 323}, 952 (2001).

46. M. V. Sazhin, V. E. Zharov, A. V. Volynkin, and T.A. Kalinina,
Mon. Not. R. Astron. Soc. {\bf 300}, 287 (1998).

47. I. I. Shapiro, Phys. Rev. Lett. {\bf 13}, 789 (1964).

48. T. V. Smirnova and V. I. Shishov, Astrophys. Space Sci. {\bf
278}, 71 (2001).

49. M. Soffel, S. A. Klioner, G. Petit, et al., Astron. J. {\bf
126}, 2687 (2003).

50. L. Spitzer, Jr., Dynamical Evolution of Globular Clusters
(Princeton Univ. Press, Princeton, 1987; Mir,Moscow, 1990).

51. J. H. Taylor, Proc. IEEE {\bf 79}, 1054 (1991).

52. J. H. Taylor, R. N. Manchester, and A. G. Lyne, Astrophys. J.,
Suppl. Ser. {\bf 88}, 529 (1993).

53. J. H. Taylor and J. M. Weisberg, Astrophys. J. {\bf 345}, 434
(1989).

54. A. V. Zakharov and M. V. Sazhin, Phys. Usp. {\bf 41}, 945 (1998)

\vspace{30mm}

\hfill{\it Translated by V.Astakhov}

\newpage

\begin{figure}
\includegraphics[width=\textwidth,bb=35 40 580 430,clip]{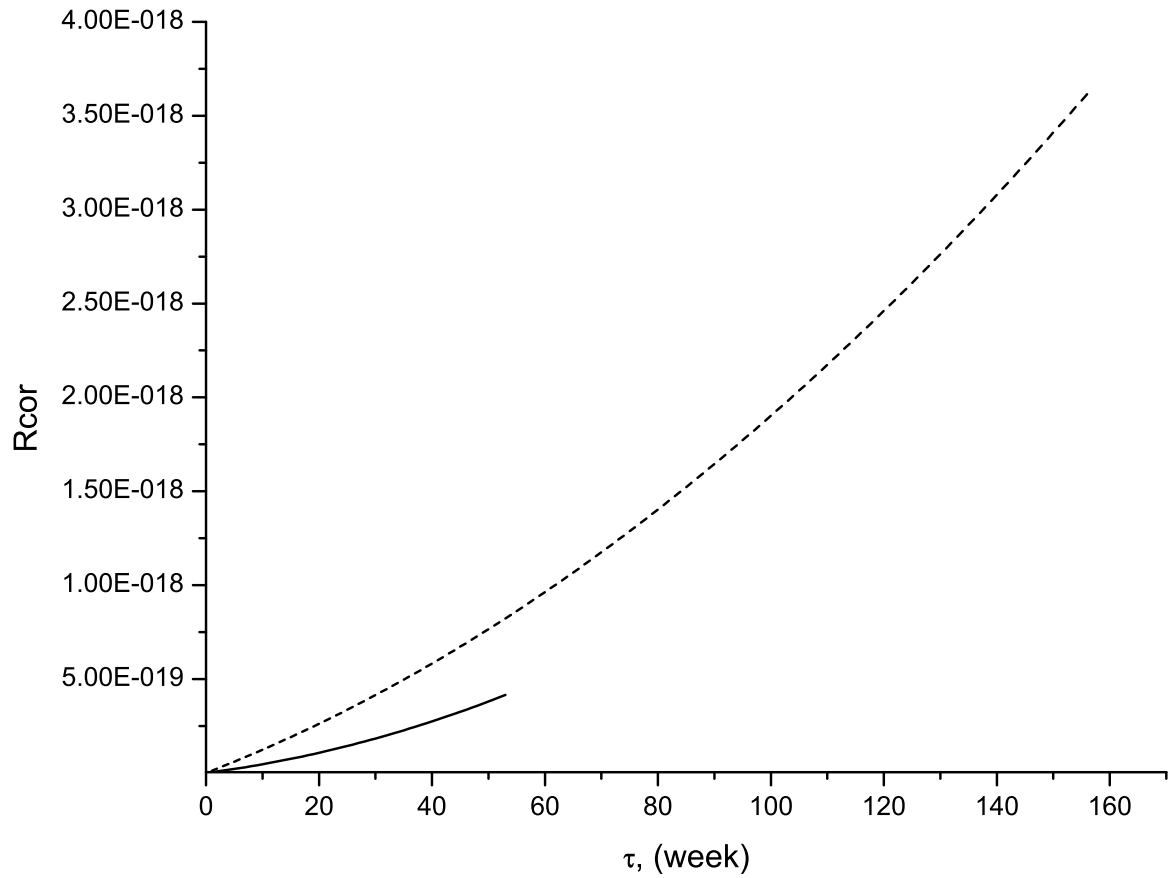}
\caption{Autocorrelation function of the stochastic Shapiro effect
for a King model with a pulsar at the center of a globular cluster
for observing times of 1 (solid line) and 3 yr (dotted line).}
\end{figure}

\begin{figure}
\includegraphics[width=\textwidth,bb=95 590 340 770,clip]{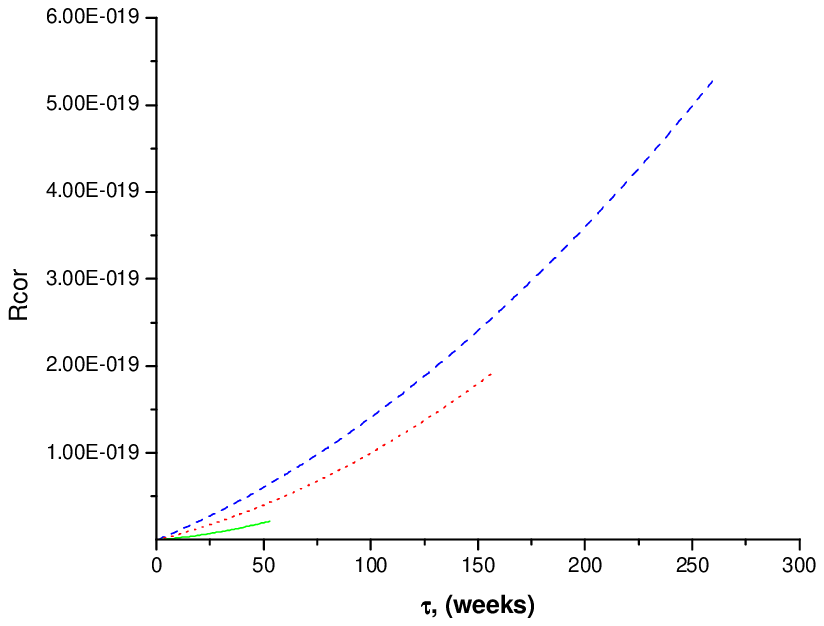}
\caption{Autocorrelation function of the stochastic Shapiro effect
for the model of an isothermal sphere with a pulsar at the cluster
center for observing times of 1 (solid line), 3 (dotted line), and 5
yr (dashed line)}
\end{figure}

\begin{figure}
\includegraphics[width=\textwidth,bb=35 40 580 530,clip]{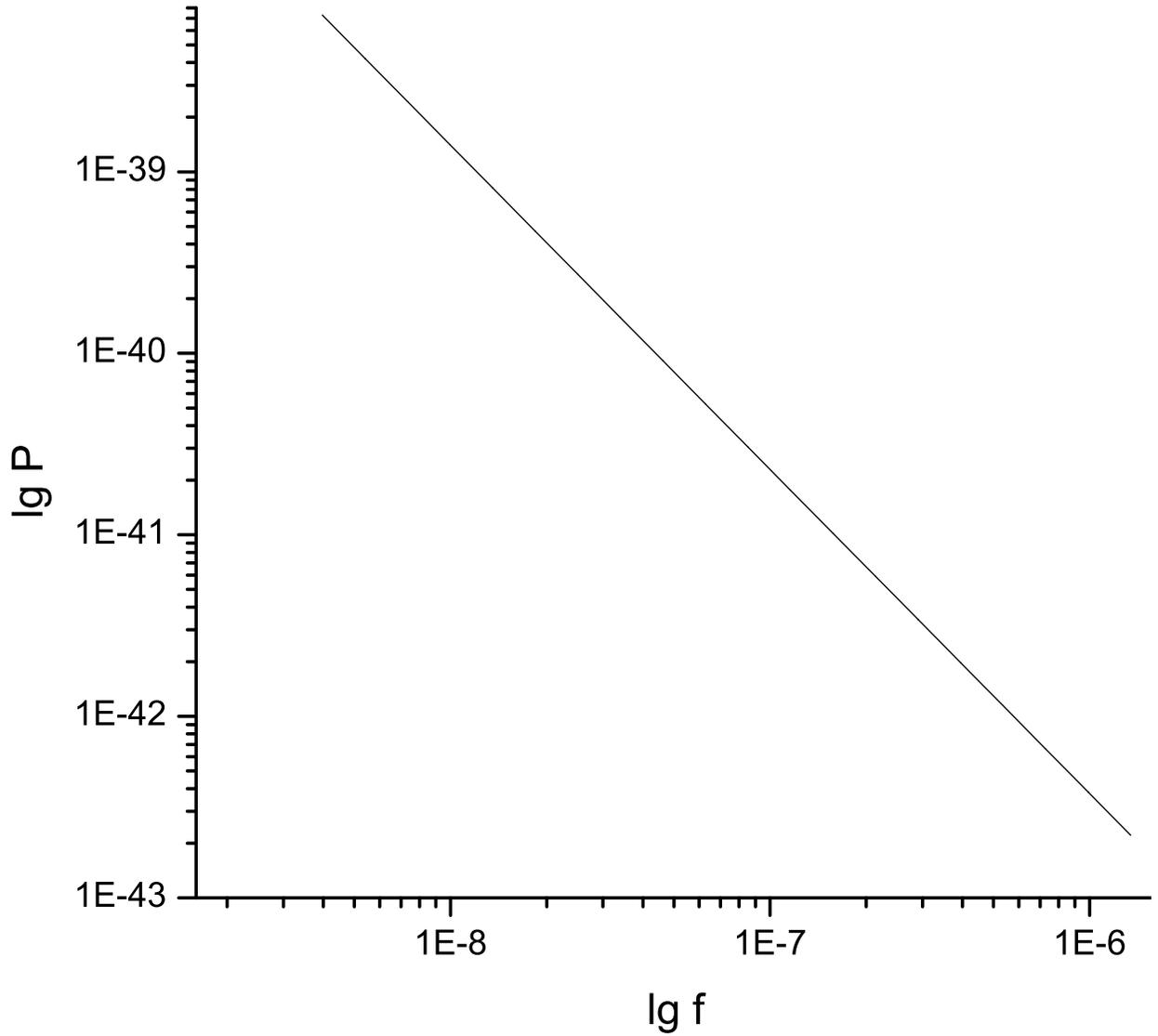}
\caption{Logarithm of the spectral power of the pulsar noise $\log
P$ versus logarithm of the Fourier frequency $\log f$ for the King
model.}
\end{figure}

\begin{figure}
\includegraphics[width=\textwidth,bb=90 580 380 780,clip]{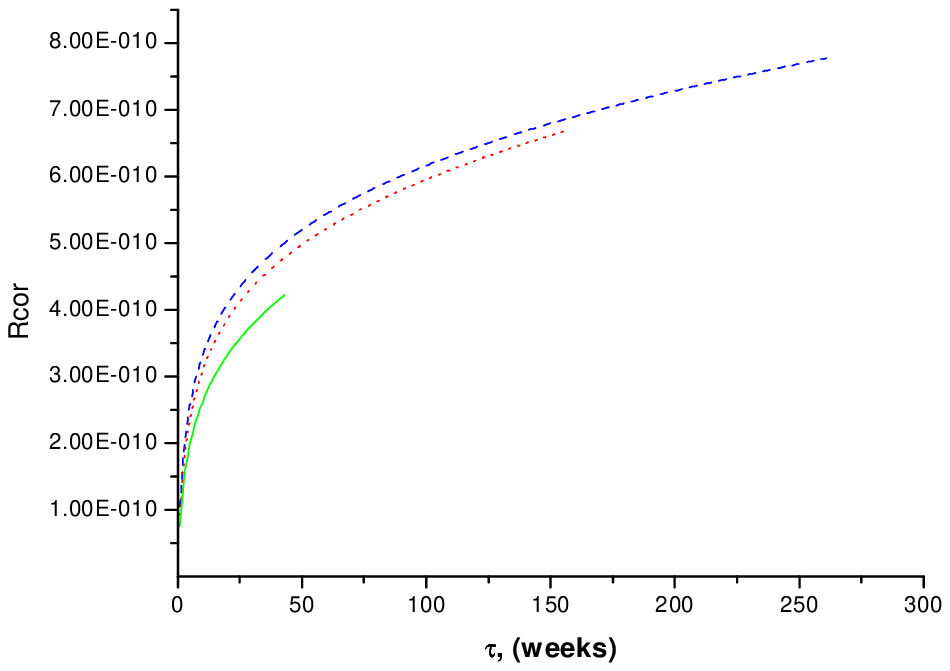}
\caption{Autocorrelation function of the stochastic Shapiro effect
for impact parameters smaller than 1 AU in the model of an
isothermal sphere with a pulsar at the center of a globular cluster
for observing times of 1 (solid line), 3 (dotted line), and 5 yr
(dashed line).}
\end{figure}

\begin{figure}
\includegraphics[width=\textwidth,bb=35 40 580 530,clip]{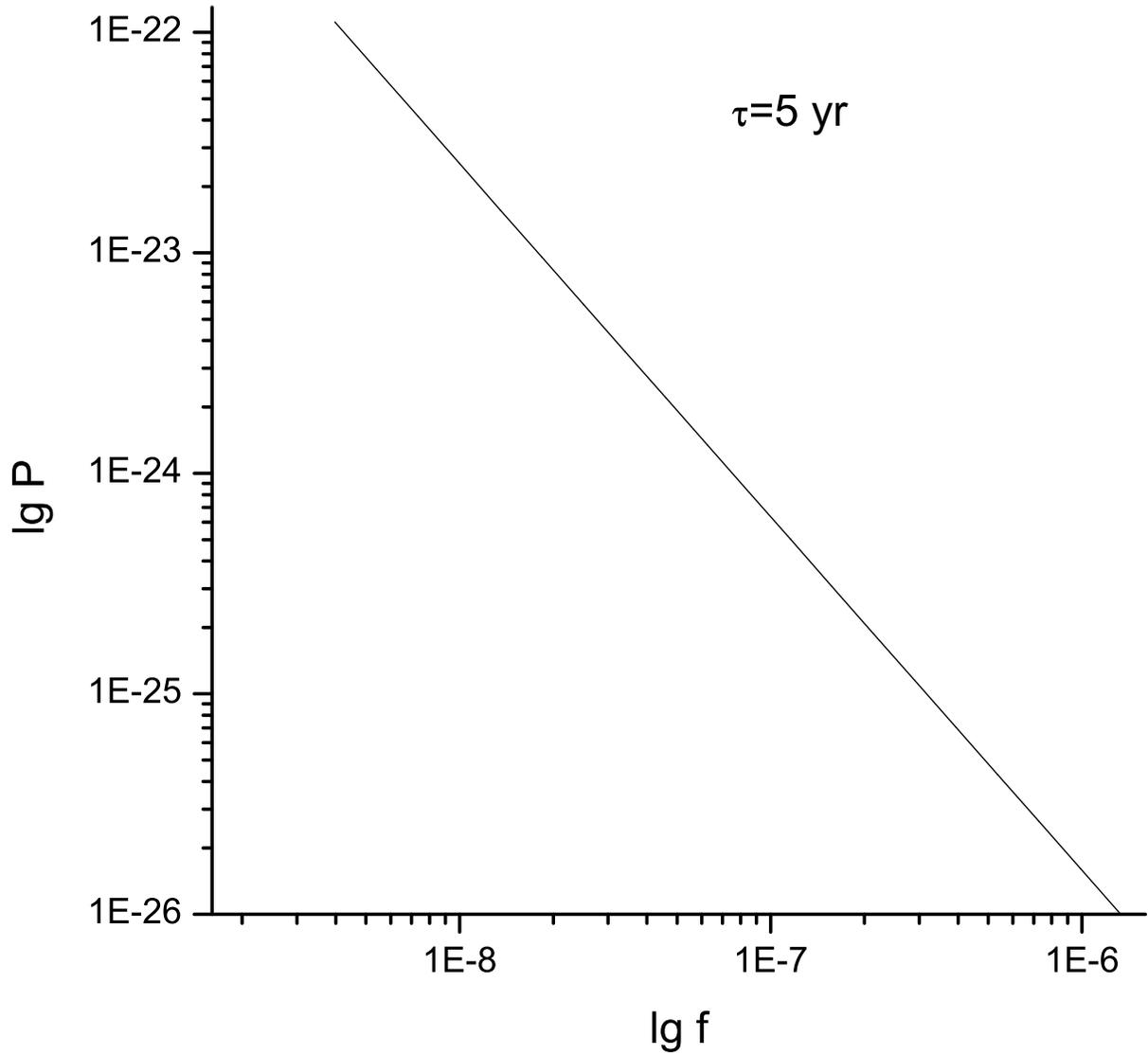}
\caption{Slope of the power spectrum for the stochastic Shapiro
effect for impact parameters smaller than 1 AU $(\tau = 5~{\rm
yr})$.}
\end{figure}

\begin{figure}
\includegraphics[width=\textwidth,bb=35 40 580 530,clip]{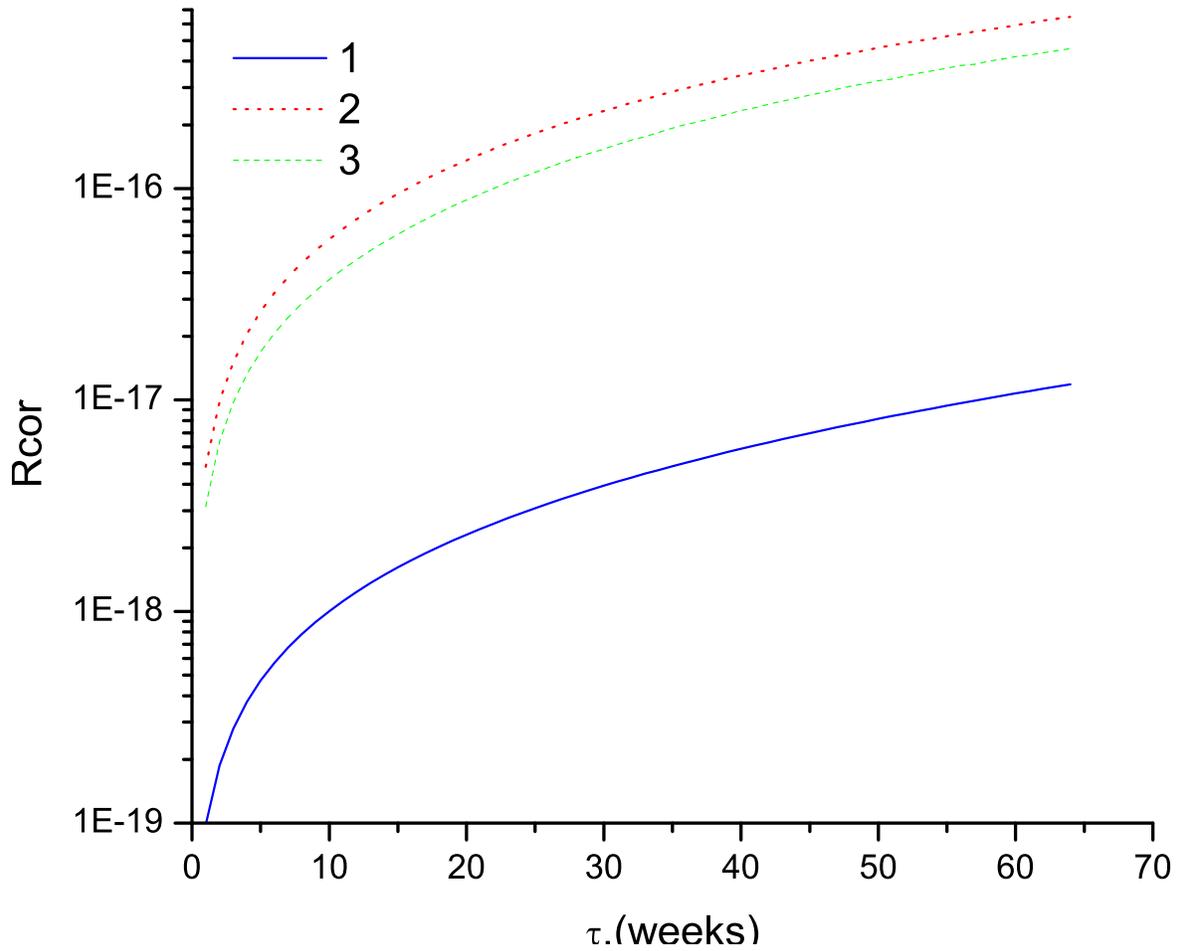}
\caption{Autocorrelation function of the stochastic Shapiro effect
for the model of an isothermal sphere with various globular cluster
core radii, 0.52 (1), 0.08 (2), and 0.1 pc (3), and with a pulsar at
the cluster center. The observing time is $\tau = 1.5$ yr.}
\end{figure}

\end{document}